\documentclass[12pt,reqno]{amsart}
\usepackage{amsfonts}
\usepackage{amssymb}
\usepackage{graphicx}
\usepackage{dsfont}
\usepackage{amsmath}
\usepackage{graphicx}
\usepackage[colorlinks=true,linkcolor=red,citecolor=blue]{hyperref}
\usepackage{mathrsfs}
\usepackage{natbib}

\begin{document}

\title{Dual $\phi$-divergences estimation in normal models}
\author{Mohamed Cherfi}
\address{Laboratoire de Statistique Th\'eorique et Appliqu\'ee (LSTA)\\ Equipe d'Accueil 3124\\ Universit\'e Pierre et Marie Curie -- Paris 6\\ Tour 15-25, 2\`eme \'etage\\ 4 place Jussieu\\  75252 Paris cedex 05 ; e-mail adresse : mohamed.cherfi@gmail.com}

\begin{abstract}
A class of robust estimators which are obtained from dual representation of $\phi$-divergences, are studied
empirically for the normal location model. Members of this class of estimators are
compared, and it is found that they are efficient at the true model and offer  an attractive alternative
to the maximum likelihood, in term of robustness .\\

{\bf Key words and phrases :}
Minimum divergence estimators; Efficiency; Robustness; $M$-estimators; Influence function.
\end{abstract}
\date{}
\maketitle
\section{Introduction}
\noindent Divergences of probability distributions  are widely used in a number of theoretical and applied statistical inference, they are also of key importance in data processing problems, see \cite{Basseville2010}. The $\phi$-divergence modeling has proved to be a  flexible and provided a powerful statistical
 modeling framework in a variety of applied and theoretical contexts, see \cite{LieseVajda2006,LieseVajda1987}, \cite{Pardo2006},  
 \cite{BroniatowskiKeziou2009} and the recent monograph by  \cite{BasuShioyaPark2011}.

\noindent Recently introduced, the
minimum divergences estimation method based on a dual
representation of the divergence between probability measures, is an appealing estimation method, it avoids the use of
nonparametric density estimation and the complications related to the bandwidth
selection. The estimators are defined in an unified way for both continuous
and discrete models. They do not require any prior smoothing and
include the classical maximum likelihood estimators as a
benchmark. These estimators called ``dual
$\phi$-divergence estimators'' (D$\phi$DE's), was shown by \cite{Keziou2003} and \cite{BroniatowskiKeziou2009}, under suitable conditions, to be consistent, asymptotically normal and asymptotically full efficient at the true model.

\noindent Application of dual representation of $\phi$-divergences have been considered by many authors, we cite among others, \cite{KeziouLeoni2008} for semi-parametric two-sample density ratio models, robust tests based on saddlepoint approximations in \cite{TomaLeoni-Aubin2009},  bootstrapped $\phi$-divergences estimates are considered in \cite{BouzebdaCherfi2011}, extension of dual $\phi$-divergences estimators to right censored data are introduced in \cite{Cherfi2011}, for estimation and tests in copula models we refer to \cite{BouzebdaKeziou2010b} and the references therein.

\noindent \noindent The $\phi$-divergences estimators are motivated by the fact that a
suitable choice of the divergence may lead to an estimate more
robust than the MLE one, see \cite{JimenezShao2001}.  \cite{TomaBroniatowski2010} studied the robustness of the D$\phi$DE's through the influence function approach. However the fact that the D$\phi$DE's have unbounded influence functions in the normal location model, causes some concern that these estimators are non-robust. In this article we show that the $\epsilon$-influence functions are more suitable to capture the robustness properties of the D$\phi$DE's. We also give some insights about the choice of the escort parameter. Simulation results indicate that the  D$\phi$DE's attain the dual goal of robustness and efficiency, they are competitive with Huber $M$-estimator without the loss of efficiency at the true model.

\noindent The rest of the paper is organized as follows. In Section \ref{Background}  we provide background material concerning the dual $\phi$-divergences estimators. Section \ref{Escort}
is devoted to the choice of the escort parameter. Section \ref{robust} deals with robustness. In section \ref{example},
we illustrate the performance of the method in real data example. A simulation study described in Section \ref{Simulation}
investigates the asymptotic properties of the estimators. Section \ref{conclusion} contains some concluding remarks.

\section{Background and estimators definition}\label{Background}

\noindent The class of dual divergences estimators has been recently introduced by \cite{Keziou2003}, \cite{BroniatowskiKeziou2009}. In the following, we shortly recall their context and definition.

\noindent Recall that the $\phi$-divergence between a bounded
signed measure $Q$ and a probability $P$ on
$\mathscr{D}$, when $Q$ is absolutely continuous with
respect to $P$, is defined by
$$D_\phi(Q,P):=\int_{\mathscr{D}}
\phi\left(\frac{\mathrm{d}Q}{\mathrm{d}P}(x)\right)~\mathrm{d}P(x),$$
where $\phi$ is a convex function from $]-\infty,\infty[$ to
$[0,\infty]$ with $\phi(1)=0$.

Well-known examples of divergences are the Kullback-Leibler associated to the function $\phi(x)=x\log x-x+1$, modified Kullback-Leibler for $\phi(x)=-\log x+x-1$, the $\chi^2$-divergence with $\phi(x)=\frac{1}{2}(x-1)^2$ and the 
{H}ellinger  distance given by $\phi(x)=2(\sqrt{x}-1)^2$.
All these divergences belong to the class of the so called
``power divergences'' introduced in \cite{CressieRead1984} (see
also \cite{LieseVajda1987} chapter 2). They are defined through
the class of convex functions
\begin{equation}  \label{powerdivergence}
x\in ]0,+\infty[ \mapsto\phi_{\gamma}(x):=\frac{x^{\gamma}-\gamma
x+\gamma-1}{\gamma(\gamma-1)}
\end{equation}
if $\gamma\in\mathbb{R}\setminus \left\{0,1\right\}$,
$\phi_{0}(x):=-\log x+x-1$ and $\phi_{1}(x):=x\log x-x+1$. (For
all $\gamma\in\mathbb{R}$, we define
$\phi_\gamma(0):=\lim_{x\downarrow 0}\phi_\gamma (x)$). So, the
$KL$-divergence is associated to $\phi_1$, the $KL_m$ to $\phi_0$,
the $\chi^2$ to $\phi_2$ and the
{H}ellinger distance to $\phi_{1/2}$. We refer to \cite{LieseVajda1987} for an overview on the
origin of the concept of divergences in statistics. The reader interested in other class of $\phi$-divergences can refer to the recent paper of \cite{KusMoralesVajda2008}, which propose a simple method of construction of new families of $\phi$-divergences.

\noindent Let $X_{1}, \dots, X_{n}$ be an i.i.d. sample
 with p.m. $P_{\theta_{0}}$. Consider the problem of estimating the population parameters of interest $\theta_{0}$, when the underlying identifiable model is given by $\mathcal{P}=\{P_{\theta}: \theta\in \Theta\}$ with $\Theta$ a subset of $\mathbb{R}^{d}$. We denote $\lambda$ a dominating measure for $\mathcal{P}$ and the resulting density
of any $P_{\theta}$ is denoted $p_{\theta}$.

\noindent Let $\phi$ be a function of class $\mathcal{C}^2$, strictly convex and satisfies
\begin{equation}
\int \left| \phi ^{\prime }\left( \frac{p_{\theta }(x)}{p_{\alpha
}(x)}\right) \right| ~\mathrm{d}P_{\theta }(x)<\infty .  \label{condition
integrabilite}
\end{equation}
By Lemma 3.2 in \cite{BroniatowskiKeziou2006}, if the function $\phi$ satisfies:
 There exists $0 <\eta < 1$ such that for all $c$ in $\left[1-\eta, 1 + \eta\right]$,
we can find numbers $c_1$, $c_2$, $c_3$ such that
\begin{equation}\label{EqRem1}
\phi(cx)\leq c_1\phi(x) + c_2 \left|x\right| + c_3, \textrm{ for all real } x,
\end{equation}
then the assumption (\ref{condition
integrabilite}) is satisfied whenever $D_{\phi}(P_{\theta}, P_{\alpha})$ is finite. From now on, we suppose that there exists a neighborhood  $\mathcal{U}$ of $\theta_0$  for which  $D_{\phi}(P_{\theta}, P_{\alpha})<\infty$  whatever $\theta$ and $\alpha$ in $\mathcal{U}$. Note that all the real convex functions $\phi_{\gamma}$ pertaining to the class of power divergences defined in (\ref{powerdivergence}) satisfy the condition (\ref{EqRem1}). 

\noindent Under the above conditions, the $\phi$-divergence:
\begin{equation*}
    D_{\phi}(P_{\theta}, P_{\theta_{0}})=\int
\phi\left(\frac{p_{\theta}}{p_{\theta_0}}\right)~\mathrm{d}P_{\theta_0},
\end{equation*}
can be represented as the following form:
\begin{equation}\label{Dualrepresentation}
 D_{\phi}(P_{\theta}, P_{\theta_{0}})=\sup_{\alpha\in
\mathcal{U}}\int h(\theta,\alpha)~\mathrm{d}P_{\theta_{0}},
\end{equation}
where $h(\theta,\alpha):x\mapsto h(\theta,\alpha,x)$ and
\begin{equation}\label{Definition-h}
h(\theta,\alpha,x):=\int \phi ^{\prime }\left( \frac{p_{\theta }}{p_{\alpha
}}\right) ~\mathrm{d}P_{\theta }-\left[ \frac{p_{\theta }(x)}{p_{\alpha
}(x)}\phi ^{\prime }\left( \frac{p_{\theta }(x)}{p_{\alpha }(x)}\right)
-\phi \left( \frac{p_{\theta } (x)}{p_{\alpha }(x)}\right) \right].
\end{equation}
Since the supremum in (\ref{Dualrepresentation}) is unique and is attained in
$\alpha=\theta_0$, independently upon the value of $\theta$, define the class of estimators of $\theta_{0}$ by
\begin{equation}\label{dualestimator}
\widehat{\alpha}_{\phi}(\theta):=\arg\sup_{\alpha\in \mathcal{U}}\int
h(\theta,\alpha)\mathrm{d}P_{n},\;\;\theta\in \Theta,
\end{equation}
where $h(\theta,\alpha)$ is the function defined in (\ref{Definition-h}). This class is called ``dual
$\phi$-divergences estimators'' (D$\phi$DE's).

\noindent The corresponding estimating equation for the unknown parameter is then given by
\begin{equation}\label{estimatingequation}
\int\frac{\partial}{\partial\alpha}h(\theta,\alpha)~\mathrm{d}P_{n}=0.
\end{equation}

Remark that the maximum likelihood estimate belongs to the
class of estimates (\ref{dualestimator}). Indeed, it is obtained when
$\phi (x)=-\log x+x-1$, that is as the dual modified
$KL$-divergence estimate. Observe that $$\int h(\theta ,\alpha ){\rm{d}}P_{n}=-\int \log
\left( \frac{p_{\theta }}{p_{\alpha }}\right){\rm{d}}P_{n}.$$
Hence keeping in mind definitions (\ref{dualestimator}), we get
\begin{equation*}
\widehat{\alpha }_{KL_{m}}(\theta )=\arg \sup_{\alpha \in \Theta }-\int \log
\left( \frac{p_{\theta }}{p_{\alpha }}\right) {\rm{d}}P_{n}=\arg \sup_{\alpha \in
\Theta }\int \log (p_{\alpha }){\rm{d}}P_{n}={MLE},
\end{equation*}
independently upon $\theta$

\noindent Formula (\ref{Dualrepresentation}) defines a family of $M$-estimators indexed by some instrumental value of the parameter $\theta$ and by the function $\phi$ defining the divergence. In the sequel we call $\theta$ the ``escort parameter'', the choice of $\theta$ appears as
a major feature in the estimation procedure, see Section \ref{Escort} below. 

\noindent Recall that an $M$-estimator of $\psi$-type is the solution of the vector equation:
\begin{equation}\label{general-m-estimator}
    \int\psi(x;\alpha)~\mathrm{d}P_n=0,
\end{equation}
where the elements of $\psi(x;\alpha)$ represent the partial derivatives of $h(\theta,\alpha,x)$ with respect to the components of $\alpha$. For more details about $M$-estimators we may refer to \cite{Huber1981},  \cite{HampelRonchettiRousseeuw1986}, \cite{MaronnaMartinYohai2006} and the references therein.

\noindent  We apply the dual representation of $\phi$-divergences (\ref{Dualrepresentation}), which we specialize to the present setting. Consider now, the normal model with density:
\begin{equation}\label{normaldensity}
    p_{\theta,\sigma}(x)=\frac{1}{\sigma\sqrt{2\pi}}\exp\left\{-\frac{1}{2}\left(\frac{x-\theta}{\sigma}\right)^2\right\},
\end{equation}
and the power divergences family (\ref{powerdivergence}).
Observe that, for $\gamma\in\mathbb{R}\setminus\left\{0,1\right\}$,
$$\frac{1}{\gamma-1}\int\left(\frac{p_{\theta,\sigma}(x)}{p_{\alpha,\widetilde{\sigma}}(x)}\right)^{\gamma-1}p_{\theta,\sigma}(x)~\mathrm{d}x=\frac{1}{\gamma-1}\frac{\widetilde{\sigma}^{\gamma}\sigma^{-(\gamma-1)}}{\sqrt{\gamma\widetilde{\sigma}^2-(\gamma-1)\sigma^2}}\exp\left\{\frac{\gamma(\gamma-1)(\theta-\alpha)^2}{2(\gamma\widetilde{\sigma}^2-(\gamma-1)\sigma^2)}\right\}.$$
Hence,
\begin{eqnarray*}
D_{\gamma}\left(P_{\theta,\sigma},P_{n}\right)&=&\sup_{\alpha,\widetilde{\sigma}}\left\{\frac{1}{\gamma-1}\frac{\widetilde{\sigma}^{\gamma}\sigma^{-(\gamma-1)}}{\sqrt{\gamma\widetilde{\sigma}^2-(\gamma-1)\sigma^2}}\exp\left\{\frac{\gamma(\gamma-1)(\theta-\alpha)^2}{2(\gamma\widetilde{\sigma}^2-(\gamma-1)\sigma^2)}\right\}\right.\\
&&-\left.\frac{1}{\gamma n}\sum_{i=1}^n\left(\frac{\widetilde{\sigma}}{\sigma}\right)^{\gamma}\exp\left\{-\frac{\gamma}{2}\left(\left(\frac{X_i-\theta}{\sigma}\right)^2-\left(\frac{X_i-\alpha}{\widetilde{\sigma}}\right)^2\right)\right\}-\frac{1}{\gamma(\gamma-1)}\right\}. \end{eqnarray*}
For $\gamma=0$,
$$D_{KL_m}\left(P_{\theta,\sigma},P_{n}\right)=\sup_{\alpha,\widetilde{\sigma}}\left\{\frac{1}{2n}\sum_{i=1}^n\left(\left(\frac{X_i-\theta}{\sigma}\right)^2-\left(\frac{X_i-\alpha}{\widetilde{\sigma}}\right)^2\right)-\log\left(\frac{\widetilde{\sigma}}{\sigma}\right)\right\}.$$
For $\gamma=1$,
\begin{eqnarray*}
D_{KL}\left(P_{\theta,\sigma},P_{n}\right)&=&\sup_{\alpha,\widetilde{\sigma}}\left\{\frac{1}{2}\left(1-\left(\frac{\widetilde{\sigma}}{\sigma}\right)^2-\left(\frac{\theta-\alpha}{\widetilde{\sigma}}\right)^2\right)-\log\left(\frac{\widetilde{\sigma}}{\sigma}\right)\right.\\
&&-\left.\frac{1}{n}\sum_{i=1}^n\left(\frac{\widetilde{\sigma}}{\sigma}\right)\exp\left\{-\frac{1}{2}\left(\left(\frac{X_i-\theta}{\sigma}\right)^2-\left(\frac{X_i-\alpha}{\widetilde{\sigma}}\right)^2\right)\right\}+1\right\}.
\end{eqnarray*}

\noindent For the normal family $P_{\theta}\equiv\mathcal{N}(\theta,1)$, with the location parameter $\theta$ and scale $\sigma=1$. It follows that, for $\gamma\in\mathbb{R}\setminus\left\{0,1\right\}$,
\begin{eqnarray*}
D_{\gamma}\left(P_{\theta},P_{n}\right)&:=&\sup_{\alpha}\int h\left(\theta,\alpha\right){\rm{d}}P_{n}\\
&=&\sup_{\alpha}\left\{\frac{1}{\gamma-1}\exp\left\{\frac{\gamma(\gamma-1)(\theta-\alpha)^2}{2}\right\}\right.\\
&&\left.-\frac{1}{\gamma n}\sum_{i=1}^n\exp\left\{-\frac{\gamma}{2}(\theta-\alpha)(\theta+\alpha-2X_i)\right\}-\frac{1}{\gamma(\gamma-1)}\right\}. \end{eqnarray*}
For $\gamma=0$,
\begin{eqnarray*}
D_{KL_m}\left(P_{\theta},P_{n}\right)&:=&\sup_{\alpha}\int h\left(\theta,\alpha\right){\rm{d}}P_{n}\\
&=&\sup_{\alpha}\left\{\frac{1}{2n}\sum_{i=1}^n\left(\theta-\alpha\right)\left(\theta+\alpha-2X_i\right)\right\}.
\end{eqnarray*}
For $\gamma=1$,
\begin{eqnarray*}
D_{KL}\left(P_{\theta},P_{n}\right)&:=&\sup_{\alpha}\int h\left(\theta,\alpha\right){\rm{d}}P_{n}\\
&=&\sup_{\alpha}\left\{-\frac{1}{2}\left(\theta-\alpha\right)^2-\frac{1}{n}\sum_{i=1}^n\exp\left\{-\frac{1}{2}\left(\theta-\alpha\right)\left(\theta+\alpha-2X_i\right)\right\}+1\right\}.
\end{eqnarray*}
We remark that the above optimizations are a feasible computationally closed-form expressions and can be performed by any standard non linear optimization code.
\section{On the choice of the escort parameter}\label{Escort}
\begin{figure}[!ht]
\begin{center}
\centerline{\includegraphics[width=8cm]{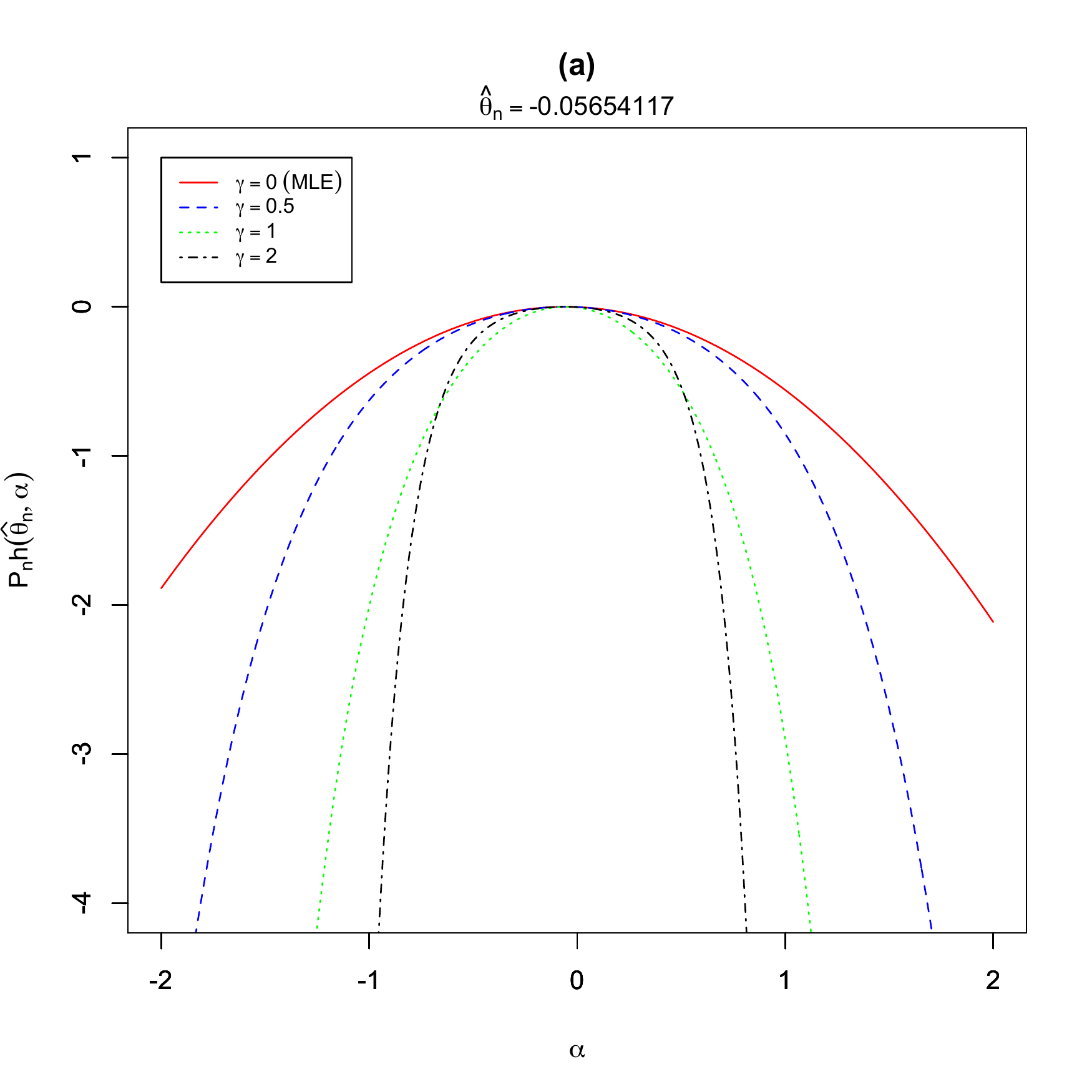}
\includegraphics[width=8cm]{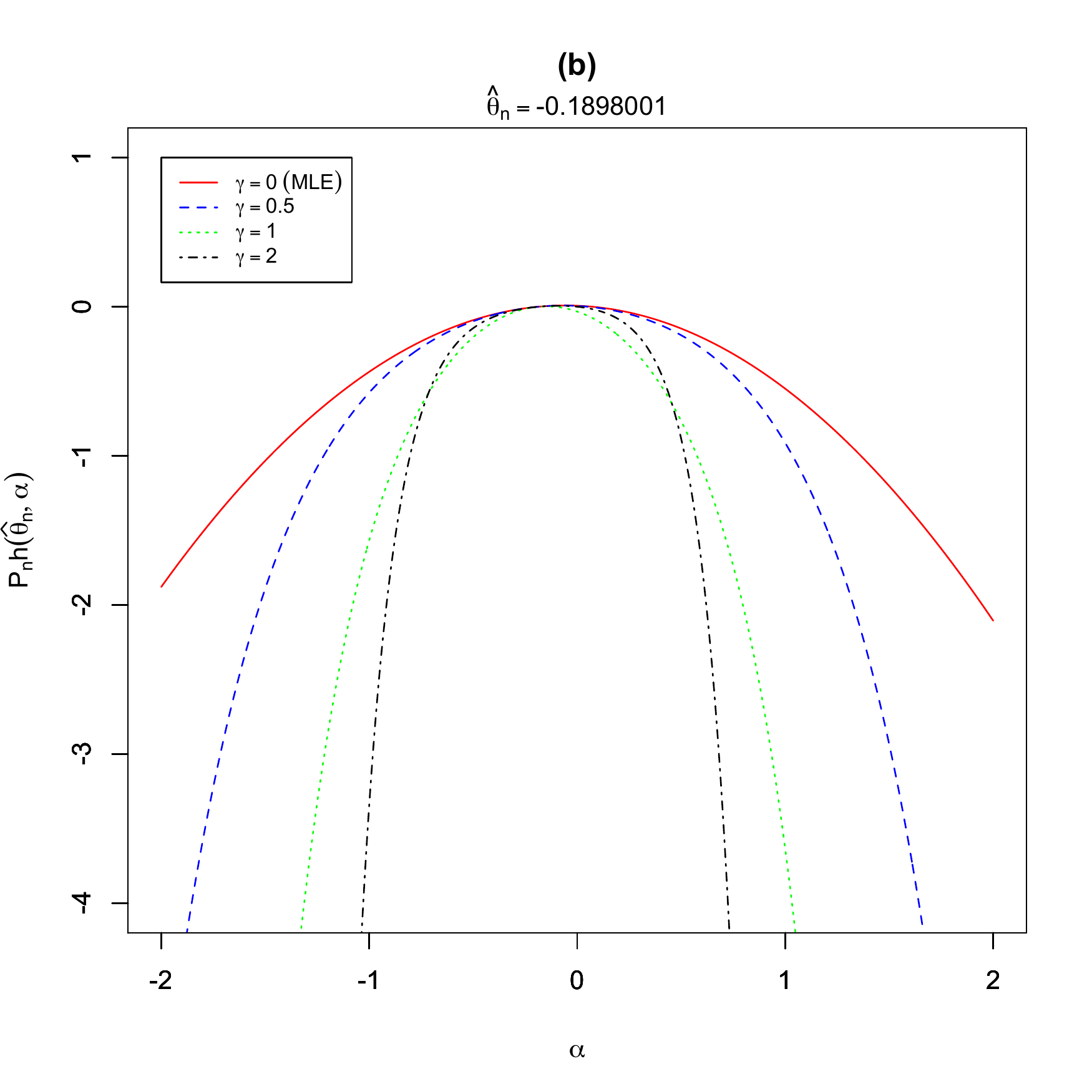}}
\end{center}
\caption{Criterion for the normal location model.} \label{normallocation1}
\end{figure}

\noindent The very peculiar choice of the escort parameter defined through $\theta=\theta_0$ has same limit properties as the MLE. The D$\phi$DE $\widehat{\alpha}_\phi\left(\theta_{0}\right)$, in this case, has variance which indeed coincides with the MLE, see for instance \cite[Theorem 2.2, (1) (b)]{Keziou2003}. This result is of some relevance, since it leaves open the choice of the divergence, while keeping good asymptotic properties.
\noindent For data generated from the distribution $\mathcal{N}(0,1)$, Figure \ref{normallocation1} shows that the global maximum of the empirical criterion $P_nh\left(\widehat{\theta}_n,\alpha\right)$ is zero, independently of the value of the escort parameter $\widehat{\theta}_n$ (the sample mean $\overline{X}$ in Figure \ref{normallocation1}(a) and the median in Figure \ref{normallocation1}(b)) for all the considered divergences.

\begin{figure}[!ht]
\begin{center}
\centerline{
\includegraphics[width=8cm]{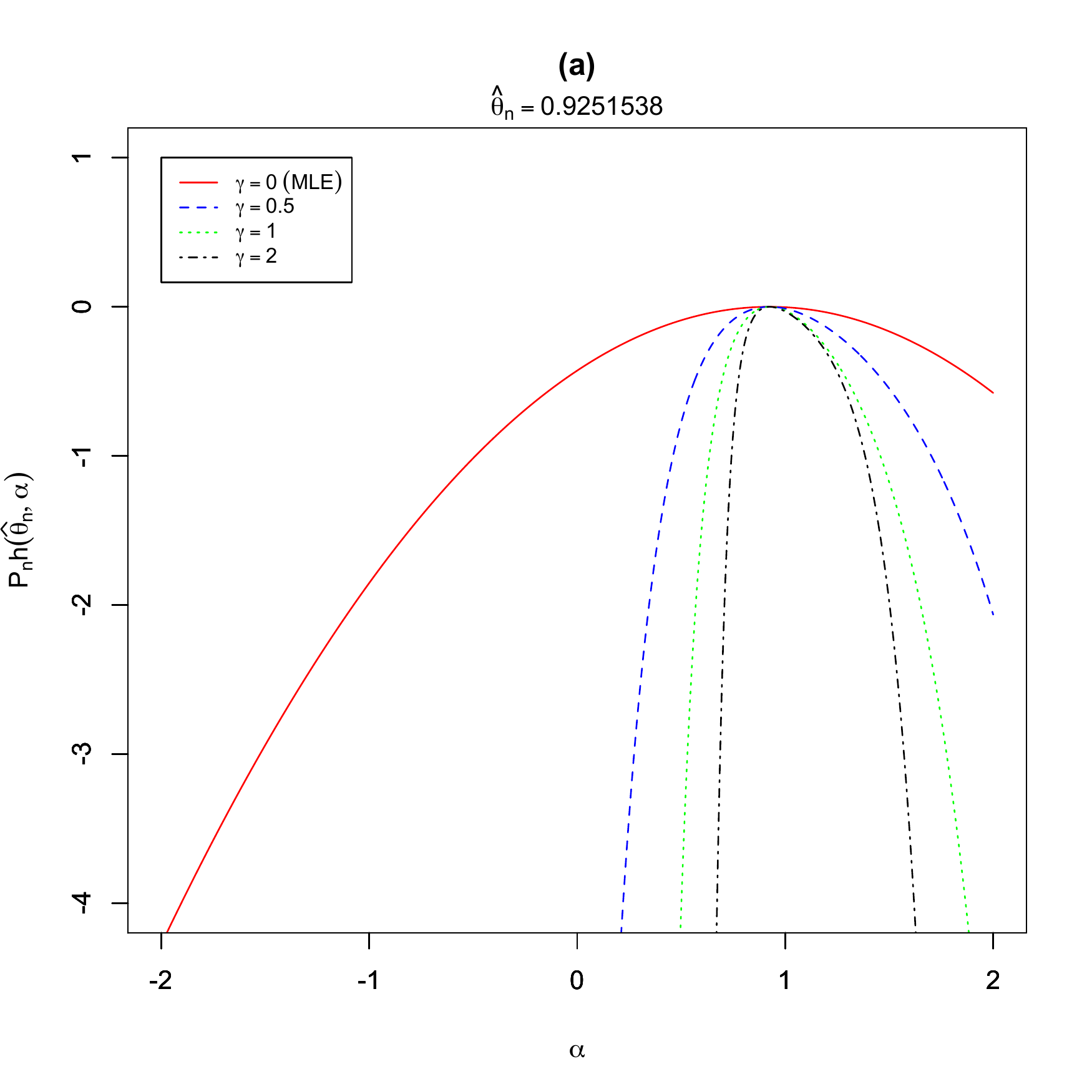}
\includegraphics[width=8cm]{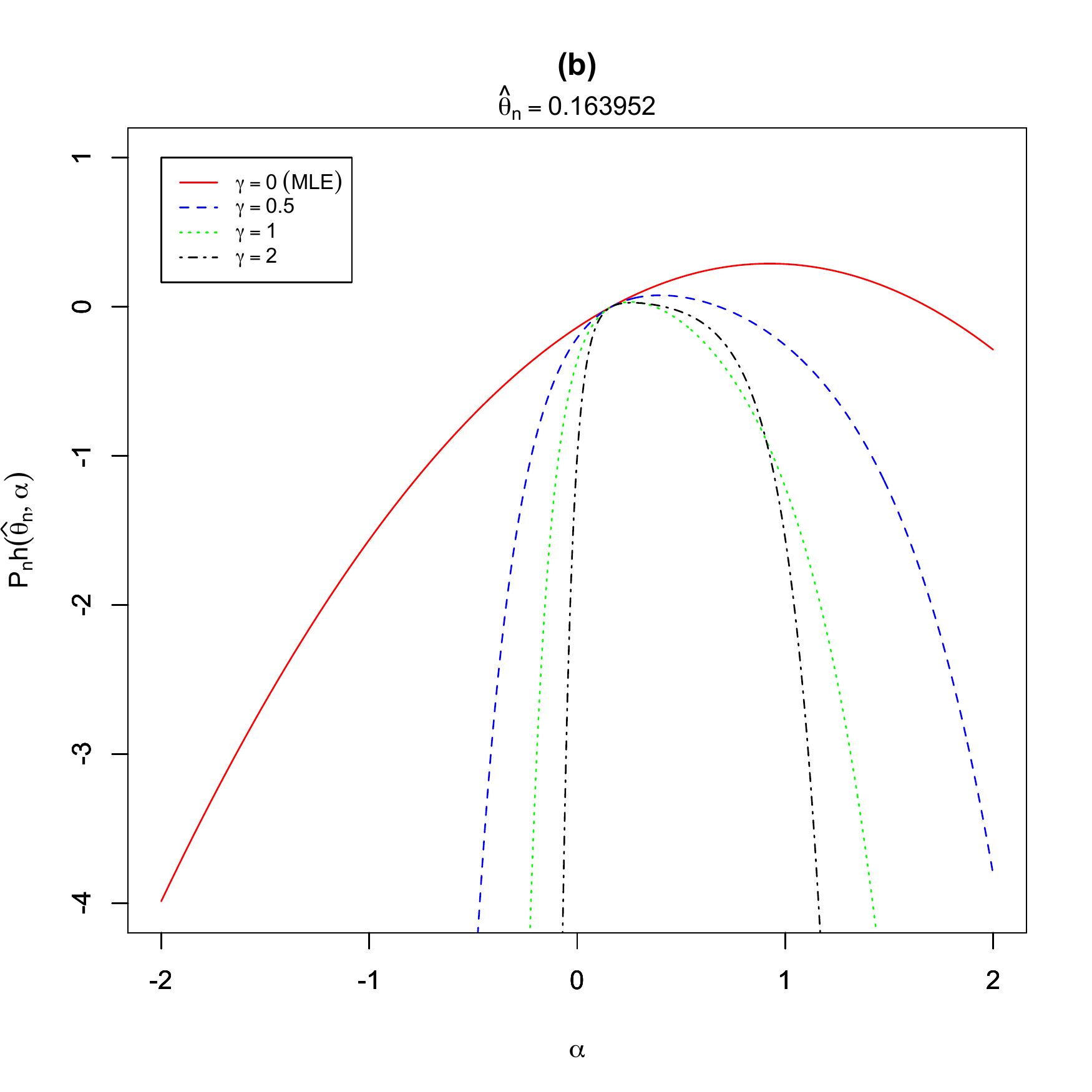}}
\end{center}
\caption{Criterion for the normal location model under contamination.} \label{normallocation2}
\end{figure}

\noindent Unlike the case of data without contamination, the choice of the escort parameter is crucial in the estimation method in the presence of outliers. We plot in Figure \ref{normallocation2} the empirical criterion $P_nh\left(\widehat{\theta}_n,\alpha\right)$, where the data are generated from the distribution
\begin{equation*}
  (1-\epsilon)\mathcal{N}(\theta_0,1)+\epsilon\delta_{10},
\end{equation*}
where $\epsilon=0.1$ and $\theta_0 = 0$.  Figure \ref{normallocation2}(a)  illustrates the empirical criterion under contamination, when we take the empirical \texttt{mean}, $\widehat{\theta}_n=\overline{X}$, as the value of the escort parameter $\theta$, it shows how the global maximum of the empirical criterion $P_nh\left(\widehat{\theta}_n,\alpha\right)$ shifts from zero to the contamination point. In Figure \ref{normallocation2}(b), the choice of the \texttt{median} as escort parameter value leads to the position of the global maximum
remains close to $ \alpha= 0$ for {H}ellinger ($\gamma=0.5$), $\chi^2$ ($\gamma=2$)  and $KL$-divergence ($\gamma=1$), while the criterion associated to the $KL_m$-divergence ($\gamma=0$, the maximum is the MLE) stills affected by the presence of outliers.

\noindent In practice, the consequence is that the escort parameter should be chosen as a robust estimator of $\theta_0$, say $\widehat{\theta}_{n}$.

\noindent Observe that for the power divergences family (\ref{powerdivergence}), the estimating equation (\ref{estimatingequation}) reduces to
\begin{equation}\label{estimatingequation-pd}
-\int
\left(\frac{p_{\theta}(x)}{p_{\alpha}(x)}\right)^{\gamma-1}\frac{\dot{p}_{\alpha}(x)}{p_{\alpha}(x)}p_{\theta}(x)~\mathrm{d}x+\frac{1}{n}\sum_{i=1}^{n}\left(\frac{p_{\theta}(X_{i})}{p_{\alpha}(X_{i})}\right)^{\gamma}\frac{\dot{p}_{\alpha}(X_{i})}{p_{\alpha}(X_{i})}=0,
\end{equation}
and the estimate $\widehat{\alpha}_\phi(\theta)$ is the solution in $\alpha$ of (\ref{estimatingequation-pd}). An improvement of the present estimate results in the plugging of a preliminary robust estimate of $\theta_0$, say $\widehat{\theta}_{n}$, as an adaptive escort parameter $\theta$ choice.

\begin{figure}[!ht]
\centerline{\includegraphics[width=8cm]{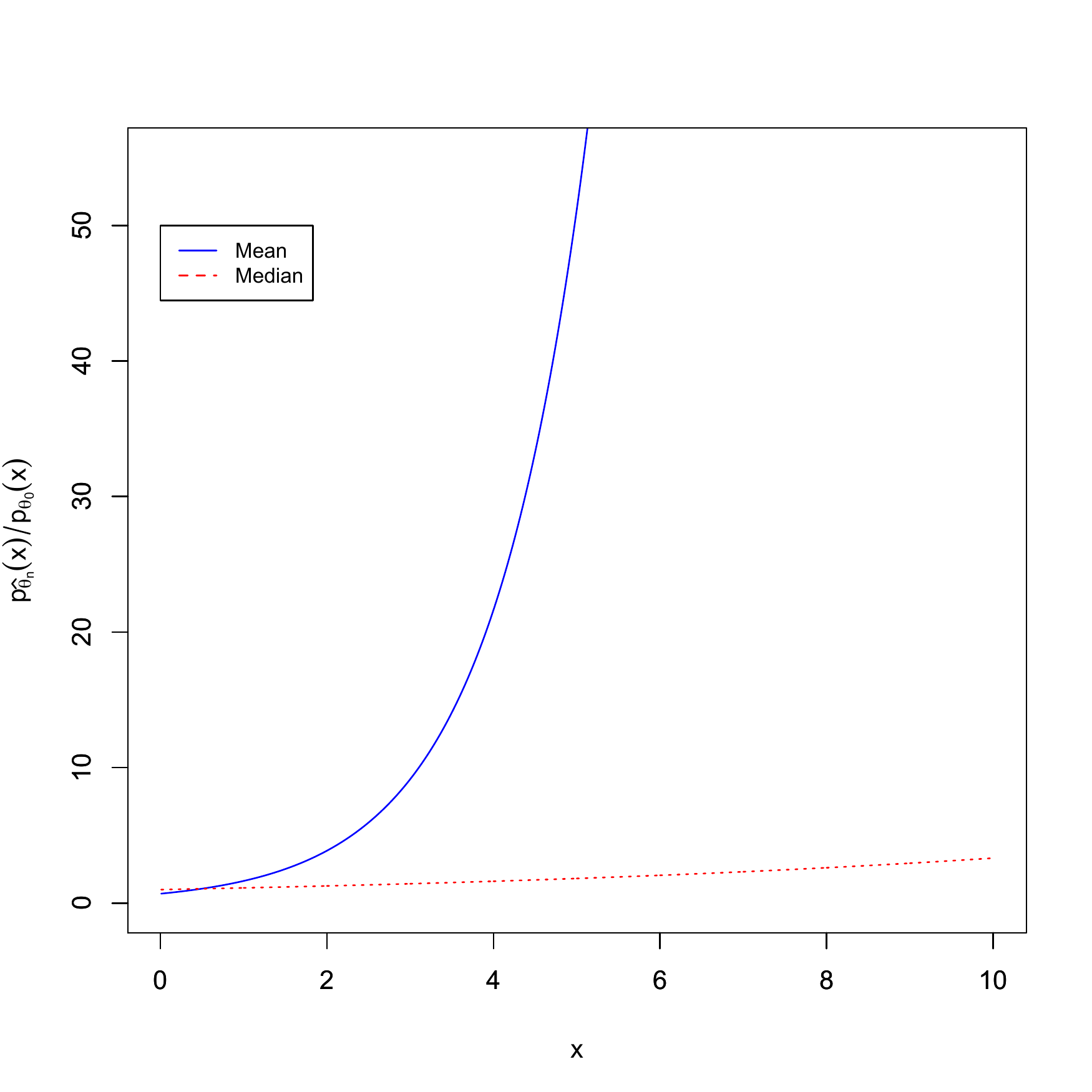}}
\caption{Behaviour of the ratio $\displaystyle{\frac{p_{\widehat{\theta}_{n}}(x)}{p_{\theta_0}(x)}}$ under contamination, for a randomly generated Normal sample $\mathcal{N}(\theta_0,1)$ of size $100$  with $10\%$ of  contamination  by $10$.} \label{ratio}
\end{figure}
\noindent Let $x$ be some outlier, the role of the outlier $x$ in (\ref{estimatingequation-pd}) appears in the term
\begin{equation}\label{leadingterm}
\left(\frac{p_{\widehat{\theta}_{n}}(x)}{p_{\alpha}(x)}\right)^{\gamma}\frac{\dot{p}_{\alpha}(x)}{p_{\alpha}(x)}.
\end{equation}
The estimate $\widehat{\alpha}_\phi(\theta)$ is robust if this term is stable. That is, if it is small when $\alpha$ is near $\theta_0$. If the escort parameter $\widehat{\theta}_{n}$ is not a robust estimator, the ratio $\displaystyle{\frac{p_{\widehat{\theta}_{n}}(x)}{p_{\theta_0}(x)}}$ can be very large, see Figure \ref{ratio}. This is due to the fact that the outlier $x$ will be more likely under $P_{\widehat{\theta}_{n}}$, that is $\widehat{\theta}_{n}$ will lead to an over evaluation of $p_{\widehat{\theta}_{n}}(x)$ with respect to the expected value under $\theta_0$, say $p_{\theta_0}(x)$.  To guard against such situations, compensate through the choice of $\gamma$, this requires further investigations.

\section{Robustness}\label{robust}
\noindent One method of assessing the robustness of an estimator is by considering its
influence function (IF), with the usual interpretation being that a
robust estimator will have a bounded influence function. However this requirement makes the
corresponding estimator deficient at the model in relation to the maximum
likelihood estimator which have unbounded influence function for most common models.

\noindent As we will see in the following, our class of dual $\phi$-divergences estimators have strong robustness features in spite of having the same influence
function as the maximum likelihood estimator. On the other hand, being equal to the influence function of the MLE, the influence function
of the D$\phi$DE's is potentially unbounded. Thus the robustness of the D$\phi$DE's cannot be
described through the traditional bounded influence approach. \cite{Hampel1974} claims that the use of the $\epsilon$-IF (before the limit) is preferable to
the use of the influence function to assess estimator robustness. The limiting form is
often used because it is usually easier to evaluate, and it does not depend on $\epsilon$. \cite{Beran1977} claims that for
evaluating the robustness of a functional with respect to a gross-error model, one should consider
the $\epsilon$-influence function instead of the influence function, unless the former converges
to the latter uniformly. For more details, see also Section 4.7 of \cite{BasuShioyaPark2011}.

\noindent In this section we present the $\epsilon$-influence function
technique and show that there is no intrinsic conflict between the robustness of our estimators and optimal model
efficiency.

\noindent The statistical functional associated with the estimate $\widehat{\alpha}_\phi(\theta)$ is given by
\begin{equation}\label{functional}
    T_{\theta}(P_{\theta_0})=\arg \sup_{\alpha \in \Theta }\int h(\theta,\alpha)~\mathrm{d}P_{\theta_0},
\end{equation}
which is Fisher consistent, namely $T_{\theta}(P_{\alpha})=\alpha$ for all $\alpha\in\Theta$, keeping in mind (\ref{dualestimator}).

\noindent  The influence function of the functional $T_{\theta}$ is defined
as
$$
\mathrm{IF}(x;T,P_{\theta_0})=\lim_{\epsilon\rightarrow 0}\frac{T_{\theta}\left((1-\epsilon)P_{\theta_0}+\epsilon\delta_{x}\right)-T_{\theta}\left(P_{\theta_0}\right)}{\epsilon},
$$
in which $\delta_{x}$ is the Dirac measure at point $x$. The $\epsilon$-influence function is the quotient
$$
\epsilon-\mathrm{IF}(x)=\frac{T_{\theta}\left((1-\epsilon)P_{\theta_0}+\epsilon\delta_{x}\right)-\theta_0}{\epsilon}.
$$

\noindent Using existing theory on $M$-estimators, see \cite{HampelRonchettiRousseeuw1986}, p. 230, see also Proposition 1 in \cite{TomaBroniatowski2010}, the influence function of the functional $T_{\theta}$ corresponding to an estimator
$\widehat{\alpha}_{\phi}(\theta)$ is given by
\begin{eqnarray}\label{defInfluence}
\nonumber{\rm{IF}}(x;T_{\theta},P_{\theta_0})&:=&-S^{-1}\bigg\{\int\phi''\left(\frac{p_{\theta}}{p_{\theta_0}}\right)\frac{p_{\theta}}{p^2_{\theta_0}}\dot{p}_{\theta_0}{\rm{d}}P_{\theta}\\
&&-\phi''\left(\frac{p_{\theta}(x)}{p_{\theta_0}(x)}\right)\frac{p^2_{\theta}(x)}{p^3_{\theta_0}(x)}\dot{p}_{\theta_0}(x)\bigg\}.
\end{eqnarray}

\noindent When $\theta=\theta_0$, it reduces to the influence function of the MLE given by
$${\rm{IF}}(x;T_{\theta_0},P_{\theta_0})=I^{-1}_{\theta_0}\frac{\dot{p}_{\theta_0}(x)}{p_{\theta_0}(x)},$$
where $I_{\theta_0}$ is the information matrix defined by
$$I_{\theta_0}=\int\frac{\dot{p}_{\theta_0}\dot{p}_{\theta_0}^\top}{p_{\theta_0}}{\rm{d}}\lambda.$$

\noindent Using (\ref{defInfluence}), the influence function for the normal location model is, for $\gamma\in\mathbb{R}\setminus\left\{0,1\right\}$,
\begin{equation*}
 {\rm{IF}}(x;T_{\theta},P_{\theta_0}):=   \frac{(x-\theta_0)\exp\left\{-\frac{\gamma}{2}(\theta-\theta_0)(\theta+\theta_0-2x)\right\}+\gamma(\theta_0-\theta)\exp\left\{\frac{\gamma(\gamma-1)(\theta-\theta_0)^2}{2}\right\}}{(1+\gamma^2(\theta-\theta_0)^2)\exp\left\{\frac{\gamma(\gamma-1)(\theta-\theta_0)^2}{2}\right\}}.
\end{equation*}
For $\gamma=0$ ($KL_m$),
\begin{equation*}
 {\rm{IF}}(x;T_{\theta},P_{\theta_0}):= x-\theta_0.
\end{equation*}
For $\gamma=1$ ($KL$),
\begin{equation*}
 {\rm{IF}}(x;T_{\theta},P_{\theta_0}):=   \frac{(x-\theta_0)\exp\left\{-\frac{1}{2}(\theta-\theta_0)(\theta+\theta_0-2x)\right\}+(\theta_0-\theta)}{1+(\theta-\theta_0)^2}.
\end{equation*}
Remark that when $\theta=\theta_0$, the influence functions of $T_{\theta}$ coincide for all $\gamma$ and are equal to the influence function of the {MLE},
\begin{equation*}
 {\rm{IF}}(x;{\rm{MLE}},P_{\theta_0}):= x-\theta_0.
\end{equation*}
 
\begin{figure}[!ht]
\begin{center}
\centerline{
\includegraphics[width=8cm]{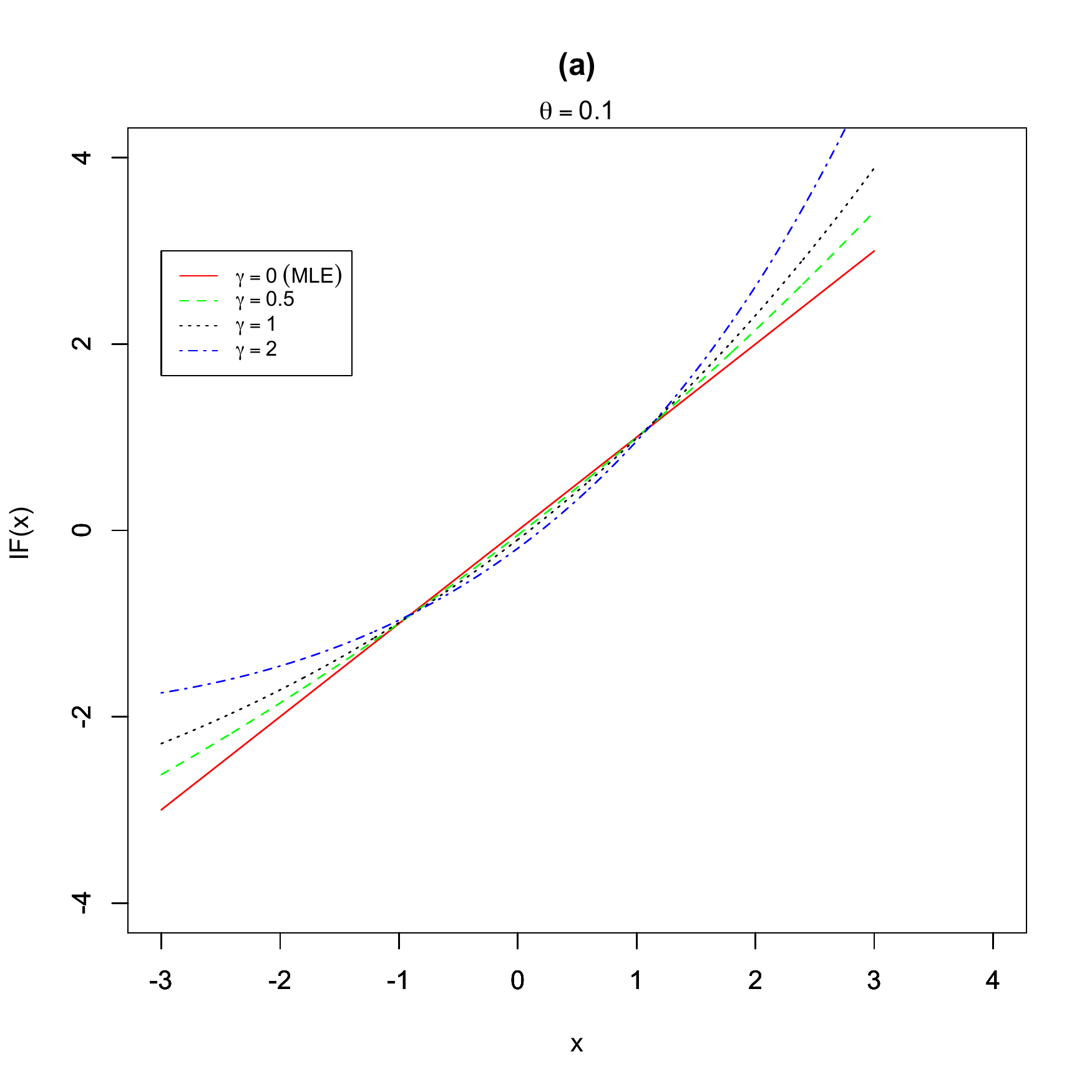}
\includegraphics[width=8cm]{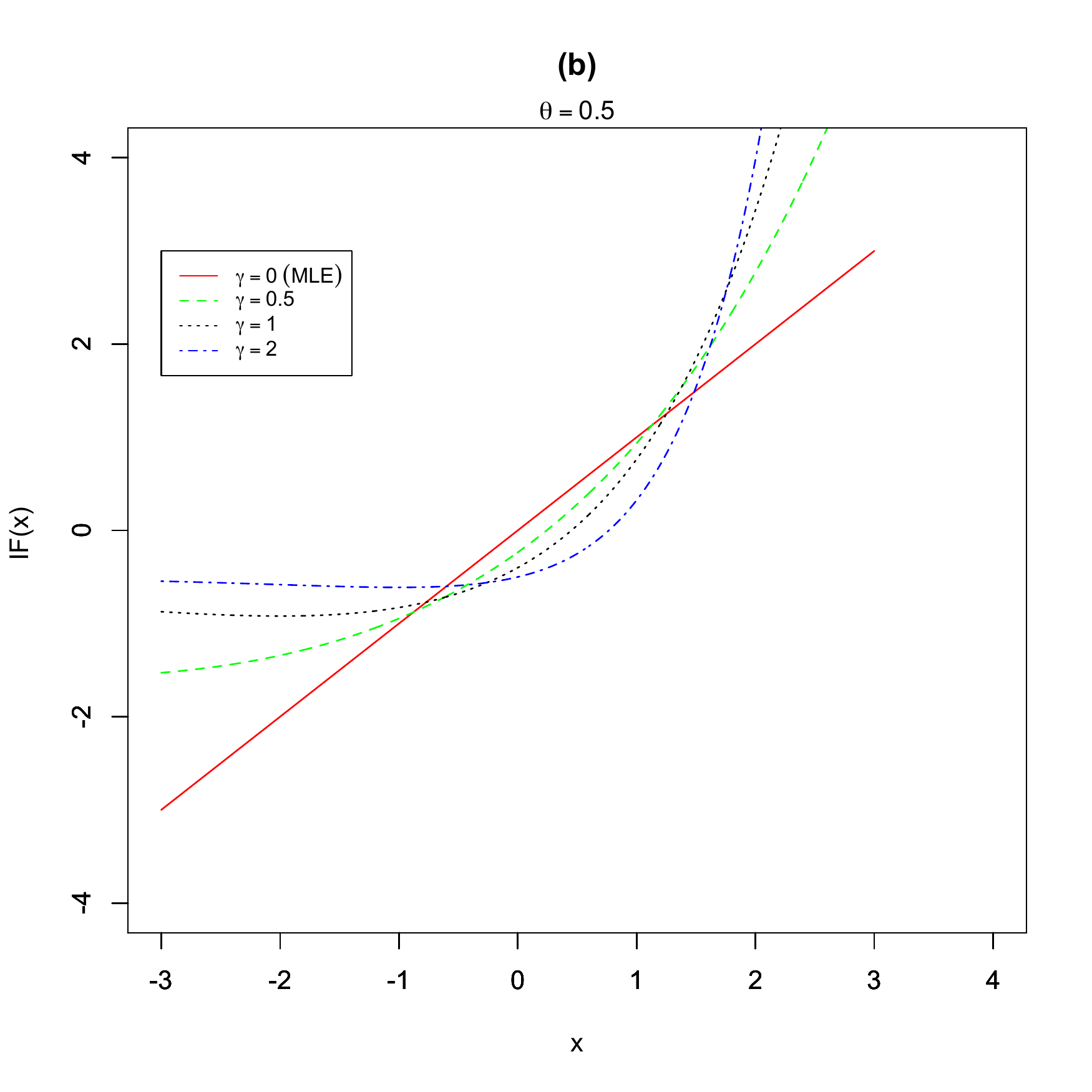}}
\end{center}
\caption{Influence functions of the D$\phi$DE's for the normal location model.} \label{Influ0}
\end{figure}
\noindent Figure \ref{Influ0}, presents the influence functions of the dual $\phi$-divergences estimators in the normal location model, with $\theta=0.1$ and $\theta=0.5$. We can see that the influence functions are unbounded. Thus, the D$\phi$DE's are examples of estimators for which the limiting form of their influence functions does not reliably provide information about the form of the $\epsilon$-IF's. 

\begin{figure}[!ht]
\begin{center}
\centerline{\includegraphics[width=8cm]{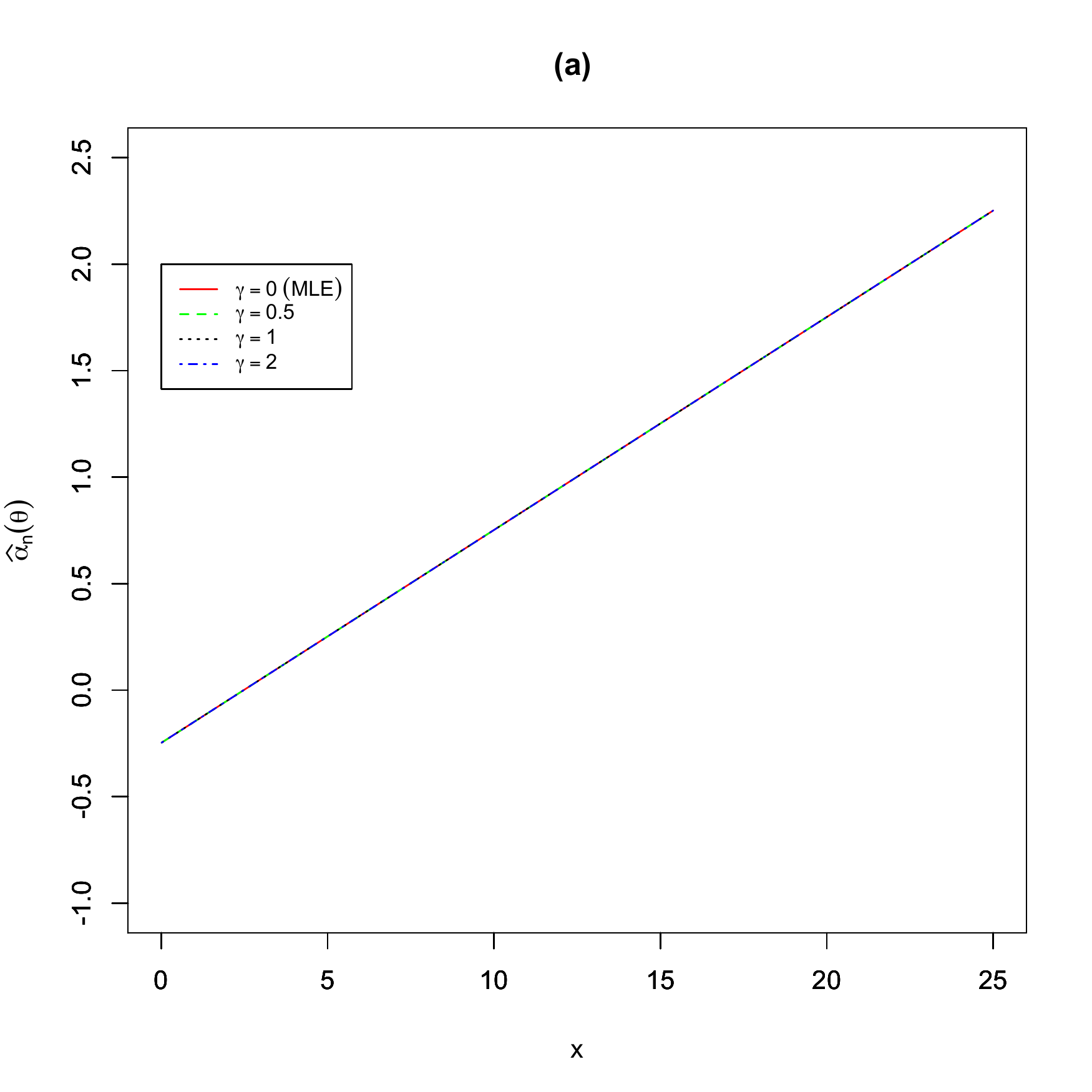}
\includegraphics[width=8cm]{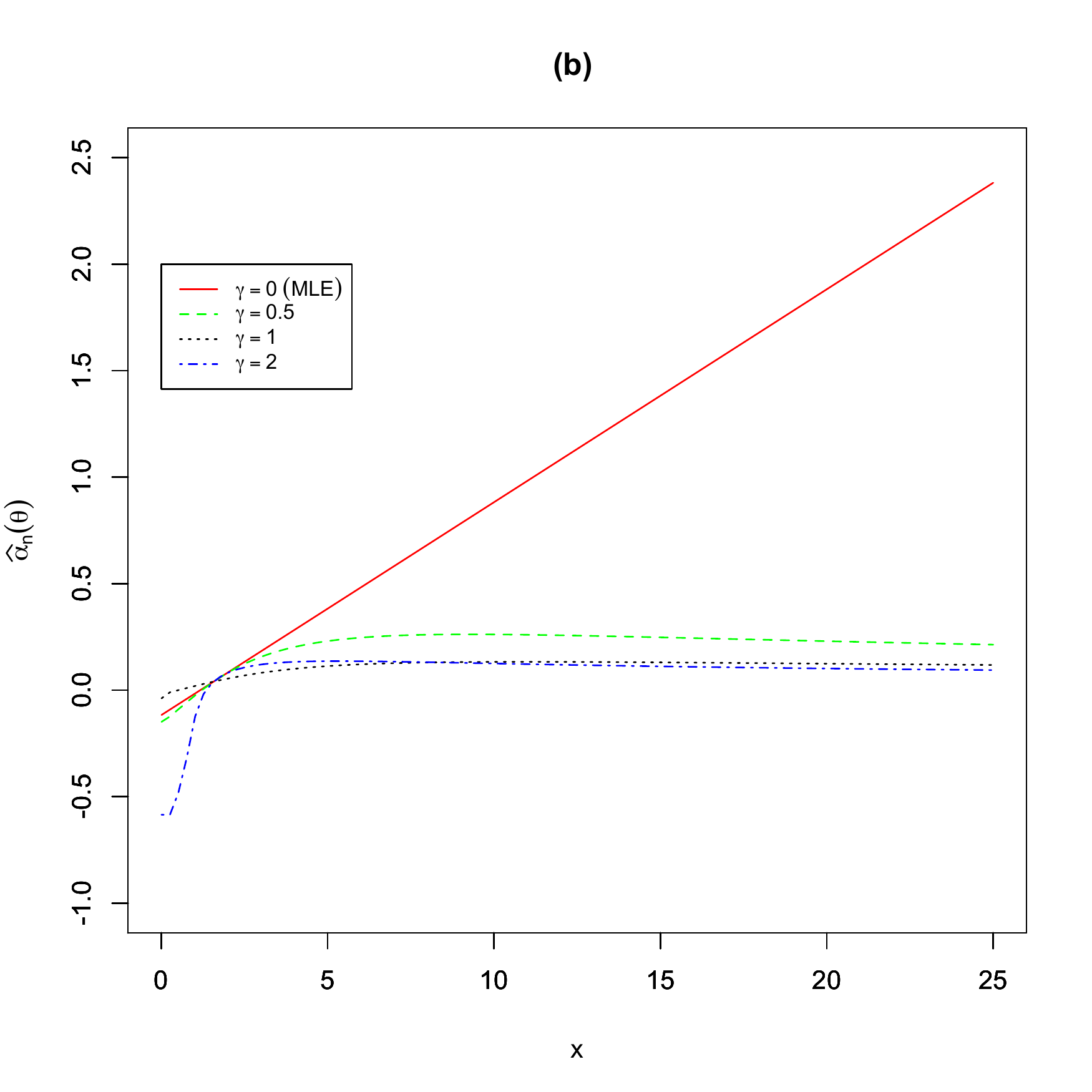}}
\end{center}
\caption{Empirical $\epsilon$-influence functions, $\epsilon=0.1$. } \label{Influ1}
\end{figure}

\begin{figure}[!ht]
\begin{center}
\centerline{\includegraphics[width=10cm]{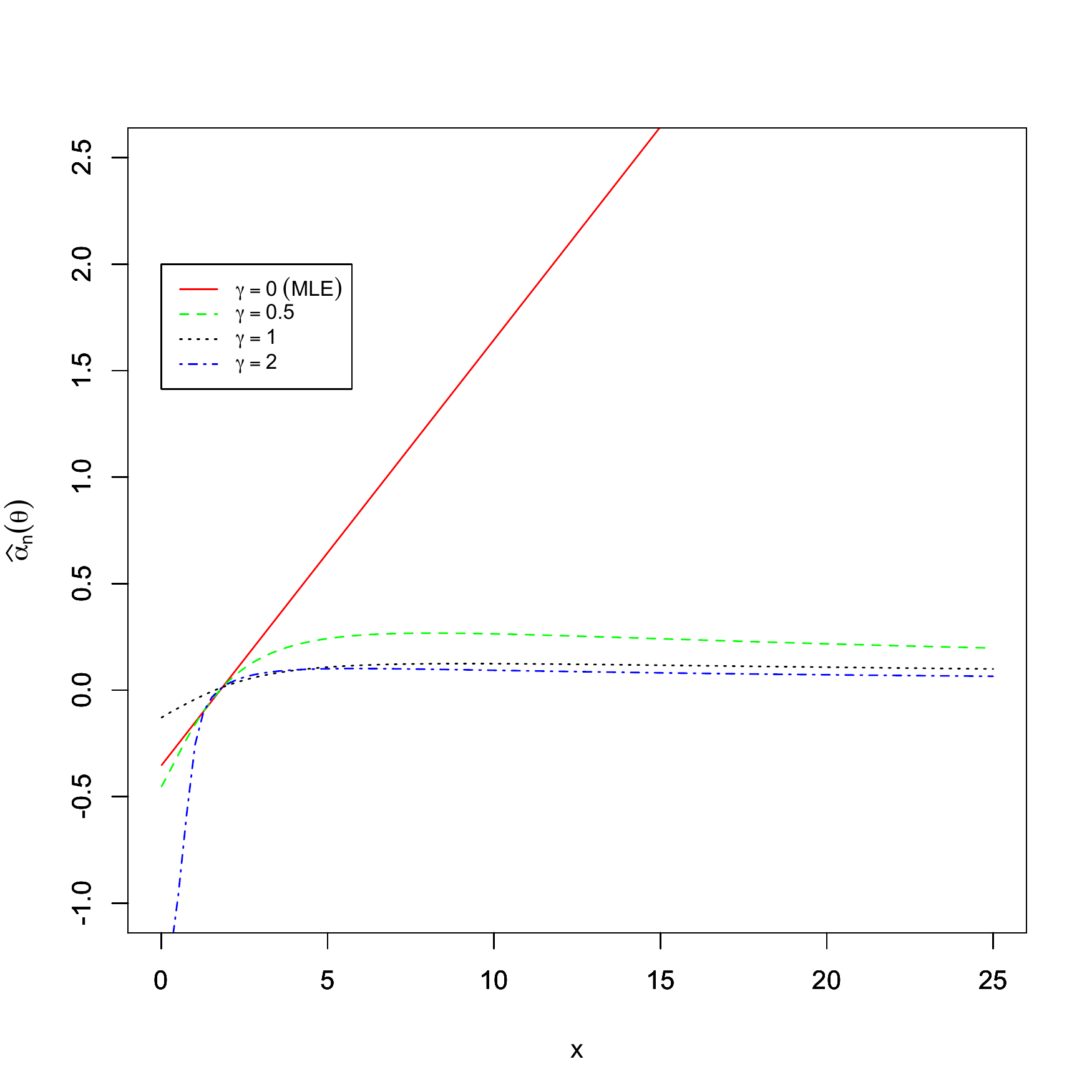}}
\end{center}
\caption{Empirical $\epsilon$-influence functions, $\epsilon=0.2$. } \label{Influ2}
\end{figure}

\noindent Empirical $\epsilon$-IF's were calculated for $\epsilon$'s$=~0.1,~0.2$. This was done by generating a sample size of $100$ for Normal distribution $\mathcal{N}(\theta_0,1)$ with $\theta_0=1$. After the data set was sorted, the largest $\epsilon n$ values were then iteratively reassigned along a grid of values ranging from $0.01$ to $25$. At each grid value the D$\phi$DE's were computed. This was repeated $1000$ times. The mean of the D$\phi$DE's at each grid were plotted against the grid. Below are the plots of the empirical average value of the D$\phi$DE's, see Figures \ref{Influ1} and \ref{Influ2}. 

Figure \ref{Influ1}(a)  illustrates the empirical $\epsilon$-IF, when we take the \texttt{mean}, $\widehat{\theta}_n=\overline{X}$, as the value of the escort parameter $\theta$, it shows that the D$\phi$DE's $\widehat{\alpha}_\phi(\theta)$ have the same empirical $\epsilon$-IF as the MLE. In Figure \ref{Influ1}(b), the choice of the \texttt{median} as escort parameter value  improve considerably the robustness properties of our estimators. 
The  D$\phi$DE's, with robust escort parameter perform very well, outliers tend have less influence on the  D$\phi$DE's. Clearly, we can see from Figure \ref{Influ2} that the MLE is greatly affected by the value of the outlier. 
\section{A Real data example}\label{example}
\noindent This example involves Newcomb’s light speed data (\cite{Stigler1977}, Table 5). The data were also analysed by \cite{BrownHwang1993}, who were trying to fit the best approximating normal distribution to the corresponding histogram. The data set shows a nice unimodal structure, and the normal
model would have provided an excellent fit to the data except for the two
large outliers.

\begin{figure}[!ht]
\begin{center}
\centerline{\includegraphics[width=10cm]{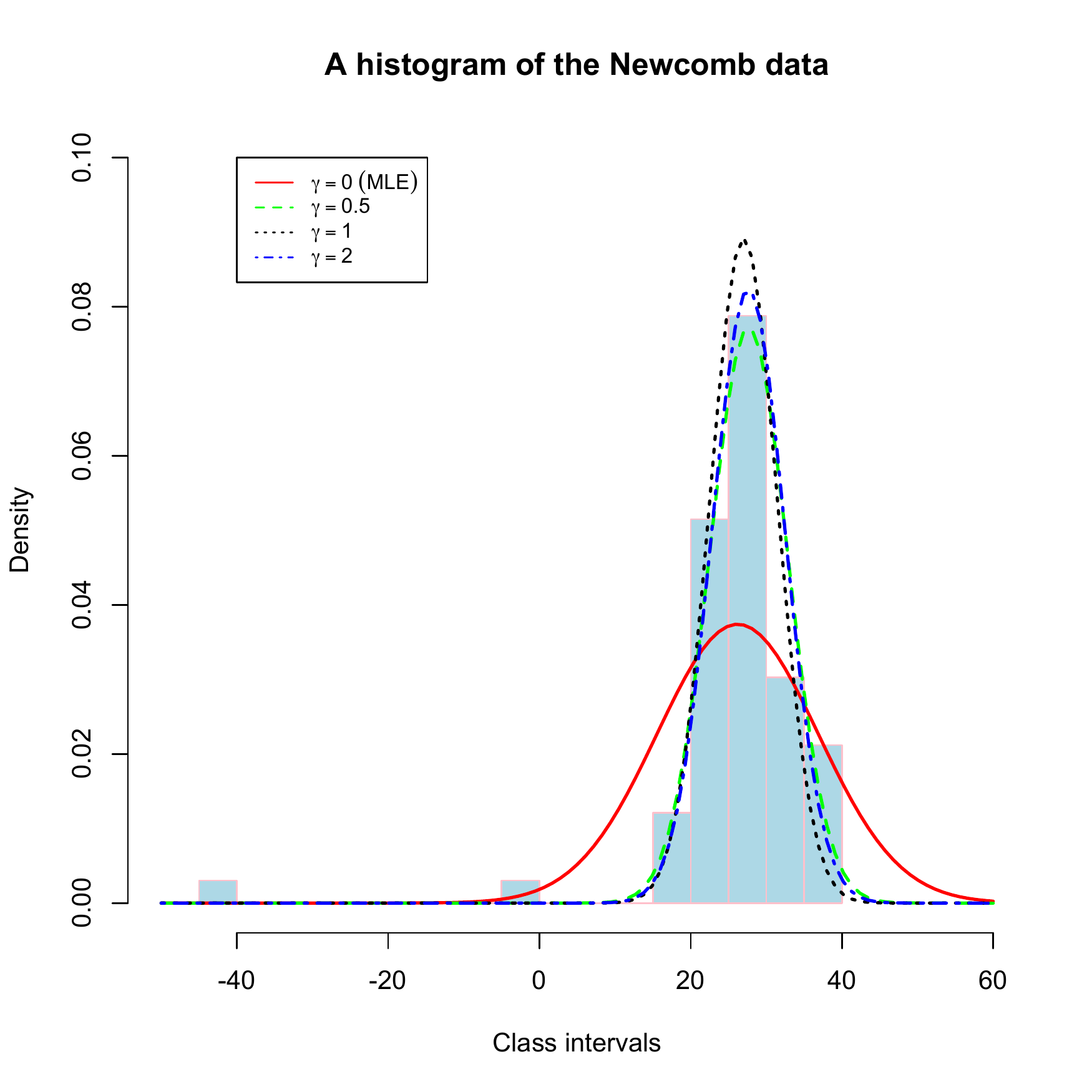}}
\end{center}
\caption{A histogram of the Newcomb data with normal fits.} \label{Hist}
\end{figure}

\noindent For the dataset, Table \ref{NewcombEstimates} gives the values of the D$\phi$DE's. These estimators
exhibit strong outlier resistance properties.  The histogram, and the normal fits using the
maximum likelihood estimate and the D$\phi$DE's are presented
in Figure \ref{Hist}, all the normal densities fit the main body of the histogram quite well, except the MLE. Note that the values of the escort parameters considered in this example are the \texttt{median} for $\theta$ and the \texttt{mad} for $\sigma$. 

\begin{table}[!ht]
\begin{center}
\begin{tabular}{lcccc}
  \hline
  & \multicolumn{4}{c}{$\gamma$} \\
  \cline{2-5}
 & 0 & 0.5 & 1 & 2 \\
  \cline{2-5} 
  $\widehat{\alpha}$  & 26.21 & 27.67 & 27.00 & 27.64 \\ 
   $\widehat{\widetilde{\sigma}}$ & 10.66 & 5.16 & 4.47 & 4.84 \\ 
   \hline
\end{tabular}
\end{center}
\caption{Estimated parameters for the Newcomb
data under the normal model.}\label{NewcombEstimates}
\end{table}

\section{Simulation}\label{Simulation}
%The performance of the D$\phi$DE's is compared with the other estimators
%described above. 
\noindent In this section, we present results of a simulation study which was conducted
to explore the properties of the D$\phi$DE's. 

\noindent A sample is generated from $\mathcal{N}(\theta_0,1)$ with  $\theta_0 = 0$. The D$\phi$DE's are calculated for samples of sizes $25,~50,~75,~100,~200$ and the hole procedure is repeated $1000$ times. In the examination of the robustness of the D$\phi$DE's, the data are generated from the distribution
\begin{equation*}
  (1-\epsilon)\mathcal{N}(\theta_0,1)+\epsilon\delta_{10},
\end{equation*}
where $\epsilon=~0.05,~0.1,~0.2$. The  value of escort parameter $\theta$ is taken to be the median.

\noindent The values of $\gamma$ are chosen to be $0,~0.5,~1,~2$ which correspond to the well known standard divergences: $KL_{m}-$divergence, the {H}ellinger distance, $KL$  and the $\chi^{2}-$divergence respectively. These estimators are also compared with some other methods, including maximum likelihood estimator (MLE) and Huber estimator (see \cite{Huber1964}) using the $\psi$-function
$$\psi(t)=\left\{\begin{array}{lcr}
t&\textrm{ if }& |t|<\tau\\
\tau{\rm sgn}(t)&\textrm{ if }& |t|\geq\tau
\end{array}\right.$$
for some constant $\tau>0$. A value of $1.4$ was used for $\tau$, for location
estimation which is in the range of values shown to perform well in the Princeton
Robustness Study (see \cite{AndrewsBickelHampelHuberRogersTukey1972}).  We carried out this analysis within the \texttt{R} Language \cite{Rcore}.

\noindent Under the true model, from Table \ref{MSE1}, as
expected,  the MLE produces most efficient estimators in this case. The D$\phi$DE's seem to be a good competitors to the MLE in terms of MSE. Recall that, theoretically, the D$\phi$DE's are asymptotically efficient. It is interesting to note that the asymptotics appear to take effect for a sample sizes of $75$ or more but the results for sample sizes below $75$ are also good, thus the  D$\phi$DE's  perform as well as the MLE at the true model.
\begin{table}
\caption{MSE of the estimates under the true model.}
\centerline{
\begin{tabular}{lcccccc}
  \hline
  & \multicolumn{6}{c}{$n$} \\
  \cline{2-7}
 & 25 & 50 & 75 & 100 & 150 & 200 \\
  \cline{2-7} 
   $\gamma$&&&&&&\\
  0 & 0.0386 & 0.0187 & 0.0129 & 0.0096 & 0.0064 & 0.0055 \\  
  0.5 & 0.0386 & 0.0187 & 0.0129 & 0.0096 & 0.0064 & 0.0055 \\  
 1 & 0.0386 & 0.0187 & 0.0129 & 0.0096 & 0.0064 & 0.0055 \\ 
  2 & 0.0440 & 0.0190 & 0.0129 & 0.0096 & 0.0064 & 0.0055 \\  
%  \hline
 $M$-Estimator & 0.0400 & 0.0199 & 0.0136 & 0.0100 & 0.0067 & 0.0057 \\  
\hline
\end{tabular}
}
\label{MSE1}
\end{table}

%\begin{table}
%\caption{MSE of the estimates under $5\%$ of contamination}
%\centerline{
%\begin{tabular}{lcccccc}
%  \hline
%  & \multicolumn{6}{c}{$n$} \\
%  \cline{2-7}
% & 25 & 50 & 75 & 100 & 150 & 200 \\
%  \cline{2-7} 
%   $\gamma$&&&&&&\\
%  0 & 0.4869 & 0.3510 & 0.3218 & 0.3022 & 0.2867 & 0.2859 \\ 
%  0.5& 0.1130 & 0.0777 & 0.0718 & 0.0680 & 0.0638 & 0.0621 \\ 
%  1 & 0.0797 & 0.0456 & 0.0390 & 0.0339 & 0.0290 & 0.0268 \\ 
%  2 & 0.0983 & 0.0483 & 0.0396 & 0.0347 & 0.0296 & 0.0273 \\ 
%%  \hline
% $M$-Estimator & 0.0735 & 0.0331 & 0.0262 & 0.0221 & 0.0183 & 0.0161 \\ 
%\hline
%\end{tabular}
%}
%\label{MSE2}
%\end{table}

\begin{table}
\caption{MSE of the estimates under $10\%$ of contamination}
\centerline{
\begin{tabular}{lcccccc}
  \hline
  & \multicolumn{6}{c}{$n$} \\
  \cline{2-7}
 & 25 & 50 & 75 & 100 & 150 & 200 \\
  \cline{2-7} 
   $\gamma$&&&&&&\\ 
  0 & 1.3999 & 1.1696 & 1.1041 & 1.0941 & 1.0505 & 1.0360 \\ 
  0.5 & 0.2004 & 0.1610 & 0.1509 & 0.1475 & 0.1447 & 0.1402 \\ 
  1 & 0.1280 & 0.0874 & 0.0780 & 0.0726 & 0.0689 & 0.0649 \\
  2 & 0.1393 & 0.0843 & 0.0731 & 0.0680 & 0.0638 & 0.0598 \\ 
%  \hline
$M$-Estimator & 0.2460 & 0.0902 & 0.0724 & 0.0646 & 0.0593 & 0.0540 \\  
\hline
\end{tabular}
}
\label{MSE3}
\end{table}

\begin{table}
\caption{MSE of the estimates under $20\%$ of contamination}
\centerline{
\begin{tabular}{lcccccc}
  \hline
  & \multicolumn{6}{c}{$n$} \\
  \cline{2-7}
 & 25 & 50 & 75 & 100 & 150 & 200 \\
  \cline{2-7} 
   $\gamma$&&&&&&\\
  0 & 4.6936 & 4.3133 & 4.1415 & 4.1344 & 4.1282 & 4.0724 \\  
  0.5 & 0.4935 & 0.4351 & 0.3947 & 0.3802 & 0.3812 & 0.3701 \\ 
  1 & 0.3253 & 0.2673 & 0.2325 & 0.2178 & 0.2163 & 0.2067 \\
 2 & 0.3092 & 0.2433 & 0.2072 & 0.1929 & 0.1906 & 0.1812 \\ 
%  \hline
 $M$-Estimator & 2.6171 & 1.5906 & 1.1710 & 0.9353 & 0.7837 & 0.6220 \\ 
\hline
\end{tabular}
}
\label{MSE4}
\end{table}

\begin{table}
\caption{MSE of the estimates under $25\%$ of contamination}
\centerline{
\begin{tabular}{lcccccc}
  \hline
  & \multicolumn{6}{c}{$n$} \\
  \cline{2-7}
 & 25 & 50 & 75 & 100 & 150 & 200 \\
  \cline{2-7} 
   $\gamma$&&&&&&\\
  0 & 7.1318 & 6.6138 & 6.5256 & 6.3594 & 6.3376 & 6.3436 \\  
  0.5 & 1.1670 & 0.6124 & 0.5920 & 0.5563 & 0.5596 & 0.5533 \\
  1 & 0.9946 & 0.3963 & 0.3754 & 0.3442 & 0.3454 & 0.3383 \\
 2 & 0.9921 & 0.3561 & 0.3347 & 0.3047 & 0.3051 & 0.2980 \\  
%  \hline
 $M$-Estimator & 5.7497 & 4.3835 & 3.9891 & 3.7579 & 3.5211 & 3.3328 \\
\hline
\end{tabular}
}
\label{MSE5}
\end{table}

\noindent We now turn to the comparison of these various estimators under contamination. We can see from Table \ref{MSE3} that for small sample sizes the D$\phi$DE's yield clearly the most robust estimates .  As $n$ increases, the $M$-Estimator obtain slightly higher performance. For high amount of contamination, we can see from Tables \ref{MSE4} and \ref{MSE5} that the D$\phi$DE with $\gamma=2$  has the smallest MSE over all other D$\phi$DE's and outperform the MLE substantially and the performance of the $M$-Estimator decreases dramatically. 

\section{Conclusions}\label{conclusion}
\noindent The aim of this paper was to investigate the new estimation procedure based on the dual representation of $\phi$-divergences for the univariate normal location model. The estimators are easily
computed and  exhibit appropriate asymptotic behavior. First, we evaluated the impact of the choice of the escort parameter on the estimates and  provided a practical choice of the escort parameter. Second, we have shown that there is no intrinsic conflict between the robustness of our estimators and optimal model
efficiency, by considering empirical $\epsilon$-influence functions.

\noindent The simulation results presented here provide solid evidence that the D$\phi$DE's in the normal location setting, obtain full efficiency at the true model while keeping good performances as robust estimators, they demonstrated unexpected empirical robustness for high amounts of contamination. Thus they provide good alternative to maximum likelihood. 

\noindent In this paper we have limited ourself to the D$\phi$DE's associated to the standard divergences, which are widely used in statistical inference. It will be
interesting to investigate theoretically the problem of the choice
of the divergence which leads to an ``\emph{optimal}'' estimate in terms of efficiency and robustness,
 we leave this study open for future research.

%\bibliographystyle{natbib}
%\bibliography{DDE}
\def\ocirc#1{\ifmmode\setbox0=\hbox{$#1$}\dimen0=\ht0 \advance\dimen0
  by1pt\rlap{\hbox to\wd0{\hss\raise\dimen0
  \hbox{\hskip.2em$\scriptscriptstyle\circ$}\hss}}#1\else {\accent"17 #1}\fi}

\end{document}